# Community Detection by ELPMeans: An Unsupervised Approach That Uses Laplacian Centrality and Clustering

Shahin Momenzadeh, Rojiar Pir Mohammadiani

Corresponding Author: Rojiar Pir Mohammadiani

**Shahin Momenzadeh, Rojiar Pir Mohammadiani**

## Abstract

Community detection in network analysis has become more intricate due to the recent hike in social networks (Cai et al., 2024). This paper suggests a new approach named ELPMeans that strives to address this challenge. For community detection in the whole network, ELPMeans combines Laplacian, Hierarchical Clustering as well as K-means algorithms. Our technique employs Laplacian centrality and minimum distance metrics for central node identification while k-means learning is used for efficient convergence to final community structure. Remarkably, ELPMeans is an unsupervised method which is not only simple to implement but also effectively tackles common problems such as random initialization of central nodes, or finding of number of communities (K). Experimental results show that our algorithm improves accuracy and reduces time complexity considerably outperforming recent approaches on real world networks. Moreover, our approach has a wide applicability range in various community detection tasks even with non-convex shapes and no prior knowledge about the number of communities present.

## Introduction

The advent of social networking sites has transformed global communication and information sharing, leading to a highly interrelated world society. These offer vast amounts of data on human interaction, giving us a different angle on society as a whole(Ali et al., 2023). Extracting meaningful insights from such immense and intricate datasets however is challenging. Social networks are better understood through community detection which involves identifying interconnected nodes that form clusters(Dey et al., 2022). Community detection aims at discovering hidden groups within a network by using its topology or node/edge attributes. The structure of communities is used to understand not only the structure but also the function of networks. They help reveal trends in social interactions, group formations and information spread. Readily availed methods for community detection include Girvan Newman and Infomap(Izem et al., 2024, Santos et al., 2024).

Nevertheless, clustering algorithms like DPC or K-means need to be included in order to improve the accuracy of community detection while allowing identification of more appropriate central nodes (Li et al., 2024, Liao et al., 2024). The choice of proper core nodes is important as far as enhanced community detection is concerned. Clustering techniques that put into consideration node relationships offer hope for a change. We want to develop a new way of detecting communities by using clustering techniques instead of traditional methods which have limitations. Our focal point is getting the right method in vectorizing nodes that preserves the network linkages. Various distance based clustering algorithms will be evaluated, and the one most suitable for community discovery will be adopted. Besides, we are going to apply it on real-world social network datasets to check if it works or not. Additionally, our method's performance depends on graph size, density and also community structure. It integrates key ideas from graph theory and clustering algorithms to come up with our proposed method ELPMeans that has been undertaken through this research work.

The main stages in ELPMeans are three as follows: initially we deploy Deep walk, machine learning algorithm transforming the network nodes into vector representations capturing structural information as well as relationships within the network. Afterward, the Laplacian centrality is computed for all nodes and a decision diagram is built to identify initial seeds of the graph and decide on the optimal number of cluster centers. Lastly, we run K-means algorithm to cluster nodes according to centers found in the previous step. The novel approach combines both graph embedding and clustering techniques, which makes it possible to effectively address community detection problem in complex networks. In this regard our findings can be useful in marketing, social network optimization or public health among others that are data driven.

This article is organized as follows: Section 2 reviews related literature on community detection and clustering algorithms by comparing their strengths and weaknesses against those of current methods employed. Subsequent section gives details about ELPMeans algorithm including its components and how they were designed. Finally, it provides experimental results explaining performance of ELPMeans on different real world datasets vis a vis other state-of-the art methods applied hereupon. Finally, section 3 concludes summarizing main contributions and proposes scope for future researches.

## Related work

The rise of social networks has caused the creation of an enormous amount of interconnected data, which makes community detection a necessary part in understanding social systems. However, such methods do not always work effectively while clustering techniques that consider node relationships are a promising alternative. There have been several documents on clustering in this area with encouraging results. It is thus clear that using clustering algorithms helps in community detection when looking at existing approaches and learning about community detection. This paper presents NAPC – a new clustering algorithm for refining density peak clustering (DPC) approach by solving three major problems. Initially, NAPC replaces Euclidean distance based similarity measure in DPC with more refined divergence distance measure. Second, it employs adjusted Boxplot theory to automate dc parameterizing or cut off distance selection process. Thirdly, NAPC also changes the way local density (q) is calculated by taking into account the new divergence distance and dc so as to provide better reflection of the data distribution as well as more accurate description of local behavior**(Yang et al., 2022)**.

This research article stresses how DPC faces challenges related to identifying centers of clusters whenever they are non-spherical ones or have different densities. These issues were addressed by the enhanced algorithm DPC-CE which selects points with greater relative distances as local centers. In this case, DPC-CE uses Euclidean distance penalty method and connectivity information as well to identify true cluster centers in the decision graph(Guo et al., 2022).

This paper presents KNN-ADPC (K-Nearest Neighbors Adaptive Density Peaks Clustering), a new algorithm for community detection in complex networks. However, here, KNN-ADPC is an improvement over traditional density peak clustering due to KNN strategy for data point assignment and adaptive merging strategy to handle over-segmented clusters. On top of it all, it provides an efficient and automated way of detecting communities, which depends on one parameter 'k'. Comparatively, experiments demonstrate that KNN-ADPC performs better than earlier techniques since it attains higher accuracy with low time complexity(Yuan et al., 2021).

The goal of this research is to improve Integrated K-means Laplacian (IKL) algorithm for real-world data clustering. The current method of producing normalized Laplacian matrix in IKL can be problematic. This study thus presents three alternative ways for constructing the normalized Laplacian matrix that address limitations inherent with the initial IKL approach. For instance, instead of Gaussian function employed in traditional IKL, 12 different kernel functions were used by scholars when developing the pairwise similarity matrix (W)(Rengasamy and Murugesan, 2022).

This study offers a new K-rank method that is intended to overcome the sensitivity of seed selection in community detection. K-rank makes use of "Rank Centrality" to identify crucial nodes in a network. K-rank improves upon community detection accuracy by choosing the initial seeds as the top-K nodes having the highest centrality values. In addition, K-rank repeatedly updates the positions of all previous seeds to optimize node assignments to communities they belong. On one hand, this paper utilizes Rank Centrality and incorporates iterative seed updates which signify an encouraging way for enhancing algorithm's performance and robustness during community detection(Jiang et al., 2013).

The main problem here is how to randomly select some of these nodes at the beginning phase, Moreover, most clustering techniques require random selection for starting point of community detection resulting into unpredictable outcomes while identifying them. There may be also difficulties in determining beforehand how many clusters are there especially in complex networks. This research introduces PC Means which is a combination of several techniques. Identifies influential nodes in local network regions and addresses the random initial node problem. determines the globally optimal number of communities in the entire network. Unsupervised and efficient, it addresses key challenges(Louafi and Titouna, 2023).

The paper introduces a novel method for clustering attributed graphs using Diverse Joint Nonnegative Matrix Tri-Factorization (DJ-NMTF). This approach simultaneously considers both the topological structure and node attributes of the graph. By employing a common factor, it integrates topological information and node features, assigning different weights to features to determine their importance in clustering. Experimental results demonstrate that this method outperforms other advanced techniques and effectively clusters graphs with diverse attributes(Mohammadi et al., 2024).

 Overall, previous research demonstrates that various clustering algorithms, including DPC, IKL, and K-rank, have significantly improved community detection in social networks and real-world data. However, some will argue that every approach has its own limitations like being sensitive to initial seed selection, computational complexity or shape and density requirements for communities. Thus, the NAPC algorithm resolves these problems by providing fresh similarity calculation methods, parameter choices and local density re-definitions. The same way DPC-CE is more accurate by incorporating distance and connectivity information. In the following chapter we will present and explore a new approach that overcomes those limitations of existing techniques to give a more complete solution for community detection.

**Methodology**

ELPMeans: A Novel Community Detection Algorithm

This article introduces a new community divide approach called ELPMeans algorithm revealing itself in another way to identify communities in graphs. There are three consecutive steps involved in this algorithm; graph embedding, center determination and clustering based on these centers. Each step depends on the one before it. Initial stage of ELPMeans algorithm involves graph embedding using deep walk that transforms nodes into vector representations. Next, Laplacian energy for all nodes is computed which followed by finding minimum distance between nodes. Finally, a decision diagram is constructed to determine the initial seeds of G and find the number of centers. The final step of the ELPMeans algorithm, called ELPMeans clustering, applies the K-means algorithm to assign nodes to the communities whose centers were determined in the previous step. Figure 1 illustrates the flowchart of the ELP Means algorithm.

---

The ELPMeans algorithm

---

**Input**: Graph, vector size, number of walks, walk length

1. Embedding graph by Deep Walk (G, vector size, num walks, walk length).
2. Calculate the distance from all nodes.
3. Calculate the minimum distance for all nodes.
4. Calculate Energy Laplacian for all nodes.
5. Find the best centers and initial centroids with the decision graph
6. Clustering (K-MEANS)
7. **Output**: Communities

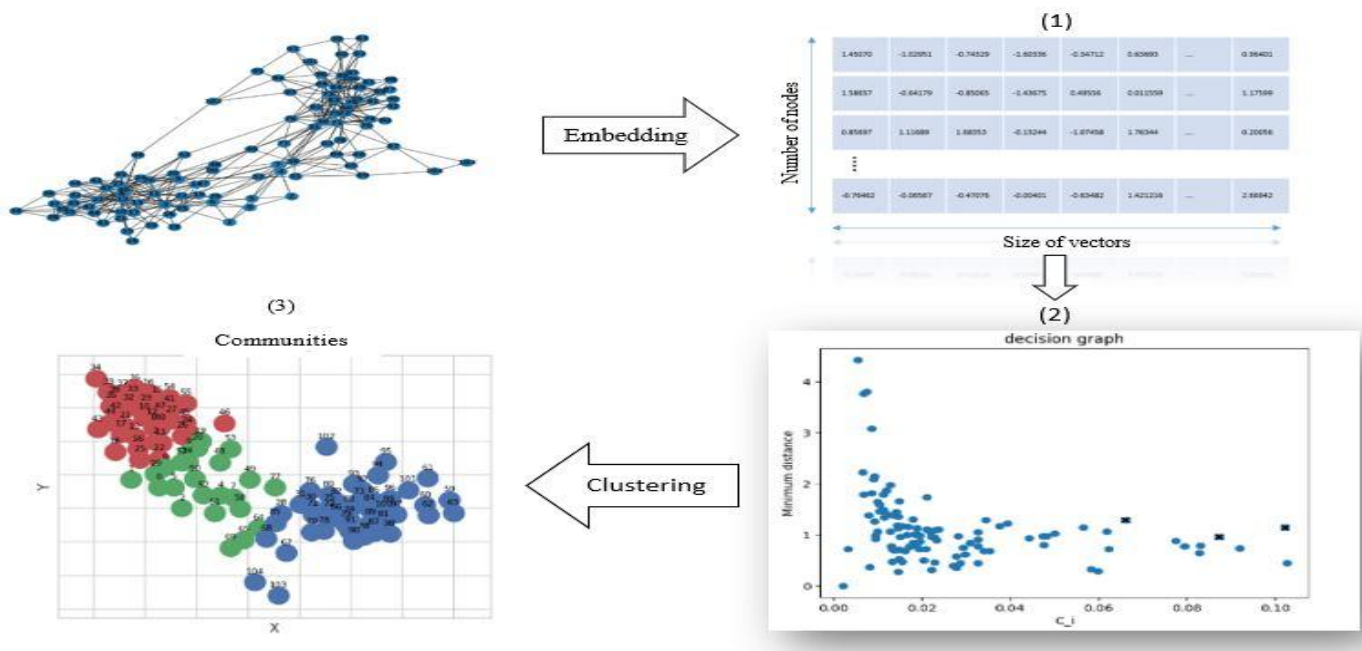

Fig. 1 ELPMeans Algorithm framework which shows three main steps: include embedding, decision graph, and detection of final communities.

The ELPC Means algorithm is applicable for community detection purposes. It can be utilized to analyze social networks, identify influential individuals within the network, and gain an understanding of the structure of complex systems. ELP Means algorithm does these steps:

1. Deep Walk is a machine learning algorithm that uses graph embedding to capture relationships within a graph(Hamedani et al., 2024). The algorithm has two main steps; first it does random walks on the graph by visiting nodes connected to each other, so as to learn the graph structure. Then, it applies the skip-gram technique which is often used in Natural Language Processing (NLP) to predict next nodes from its neighbors on these walks. Deep Walk learns meaningful embeddings for each node by refining these node vectors based on these predictions. This process effectively captures the structural information of the graph and translates it into a continuous vector space.

2. Calculation of Laplacian energy and minimum distance and drawing of decision diagram: Certain indices and notations are used here to simplify problem description. The original dataset, denoted as D = $[v_i]$, where i ranges from 1 to N, consists of arbitrary data points $v_i$. The distance between data points $v_i$ and $v_j$ is represented as $d_{ij}$. To identify cluster centers within the dataset, two indexes are assigned to each data point. The first is the local importance index $c_i$, which is based on the Laplacian centrality of data point $v_i$. The second is the minimum distance $\delta_i$ from data point $v_i$ to other nodes that possess higher Laplacian centrality. the local importance of a cluster center is higher than its neighboring nodes and there's always a large distance between a cluster center and a data point with higher local importance. Laplacian centrality of each node in the network is calculated. For a data point vi, Laplacian centrality ci is used as an evaluation index of its local importance. This data point's ci is compared with numerical value of Laplacian centrality of other data points to get set of data points with high local importance than this. Also, compute minimum distance δi between vi and each data point in this set.In case if ci and δi of vi are greater than those of any other ordinary data point, then it will be considered as the cluster center. Traversing the network, all data points with such features are recognized as cluster centers. Laplacian centrality is a method for evaluating the importance of nodes in a weighted network. It calculates the significance of a node by assessing how much the network's "Laplacian energy" changes after removing that node. This energy measure, based on the Laplacian matrix, reflects the network's overall structure and connectivity. A larger decrease in Laplacian energy upon node removal indicates a more crucial node. The study utilizes a special type of network called a "complete coupling network" to represent the data. This network is undirected, meaning connections go both ways, and it doesn't include loops where a node connects back to itself. Within this network, a weight matrix denoted by W(G) captures the strength of connections between data points. Each element $W_{i,j}$ in this matrix represents the specific weight of the edge connecting data points vi and vj. In simpler terms, the weight reflects the importance or influence of the connection between those two data points.

$$W(G) = \begin{pmatrix} 0 & W_{1,2} & \dots & W_{1,n} \\ W_{2,1} & \vdots & \ddots & W_{2,n} \\ \vdots & & & \vdots \\ W_{n,1} & W_{n,2} & \dots & 0 \end{pmatrix} \qquad (1)$$

The study creates a matrix X(G) that reflects the connectivity of each data point. It calculates the weighted sum for each data point (representing its outgoing connections) and places this sum on the corresponding diagonal entry in X(G).

$X(G) = \begin{pmatrix} X_1 & 0 & 0 & X_2 & \dots & 0 & \dots & 0 & 0 & 0 & 0 & 0 & \ddots & 0 & \dots & X_n \end{pmatrix}$   (2)   The Laplacian matrix of G: L(G) = X(G)-W(G)   (3)

$$\text{The Laplacian energy of G is:} \quad E_L(G) = \sum_{i=1}^{n} x_i^2 + 2\sum_{i<j} w_{i,j}^2 \quad (4)$$

$$\text{The Laplacian centrality of a data point } v_i \text{ is:} \quad c_i = \frac{(\Delta E)_i}{E_L(G)} = \frac{E_L(G) - E_L(G_i)}{E_L(G)} \qquad (5)$$

Where $E_L(G_i)$ is Laplacian energy after the data point $v_i$ is deleted.

**Minimum distance** $\delta_i$: Laplacian centrality ($c_i$) is used to measure the local importance of a data point ($v_i$). The ci of vi is compared to other data points, and the minimum distance ($\delta_i$) between vi and those with higher importance is calculated. This distance may reflect relevant network structure and relationships:

$$\delta_i = d_{ij} \qquad (7)$$

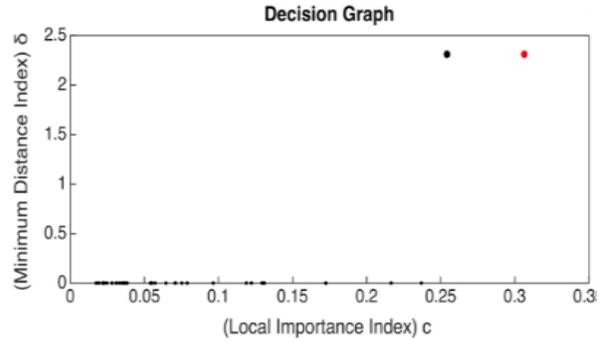

Fig. 2. Decision graph for the data points of the karate network. Each data point is shown on the decision graph according to its Laplacian centrality c and the minimum distance δ. The two-colored data points are the two cluster centers.

3. K-means for Community Detection:

While K-means is a popular clustering method that can be used to group nodes in network communities based on their similarities, it has limitations. ELPMeans overcomes these limitations by leveraging the results of a preceding step to determine the optimal number of communities (K) and identify initial centroids. This approach sidesteps the challenge of manually selecting K and has the potential to improve the final community structure. Subsequently, ELPMeans employs K-means on these predefined centroids to finalize the community assignments for each node.

**Results Experiments**

Based on the provided information, it can be concluded that the proposed method performs significantly better or similarly in detecting communities in different networks, particularly in terms of ACC and NMI metrics, compared to other methods

Accuracy (ACC): Indicates the level of overlap between the clusters created by the clustering algorithm and the true clusters of the data(Souravlas et al., 2021).

$$ACC = \frac{TP + TN}{TP + TN + FP + FN}$$

Normalized Mutual Information (NMI): Indicates the level of correlation between the clusters created by the community detection algorithm and the true clusters of the data(McDaid et al., 2011).

$$NMI = \frac{MI(X,Y)}{\sqrt{H(X) \cdot H(Y)}}$$

In this section, clustering using the proposed method with the ELPMeans algorithm has been performed. Table 4-1 presents the parameters for each dataset in the proposed method. The performance of the proposed method is superior or similar to other methods in detecting communities in the "football," "poolbook," and "karate" datasets. In the "dolphins" and "emails" datasets, the accuracy of the proposed method is comparable to the best existing methods(Zachary, 1977, Lusseau, 2003, Newman and Girvan, 2004). The proposed method relies on a parameter-sensitive deep step(Shao et al., 2023).

Table 1 Parameters for each data set in the proposed method

| Dataset | K | Vector size | Num walk | Walk length |
|---------|---|-------------|----------|-------------|
| football | 2 | 7 | 115 | 10 |
| pool book | 3 | 5 | 105 | 13 |
| Karate | 2 | 5 | 100 | 10 |
| Dolphins | 2 | 7 | 50 | 10 |
| Emails | 42 | 7 | 500 | 10 |

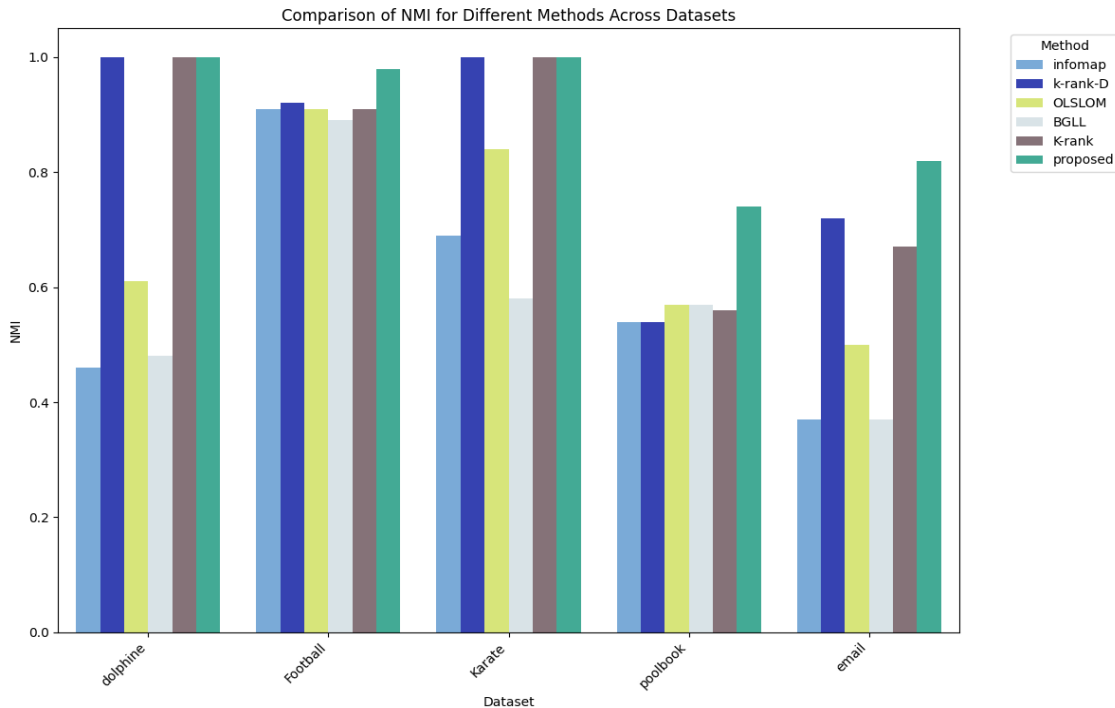

Figure 3: Comparison of the Proposed Method with Competing Algorithms based on the NMI Metric

In Figure 3, a comprehensive comparison between the proposed method and several competing algorithms, including k-rank-d, k-rank, OLSLOM, BGLL, and Infomap, is presented based on the NMI metric. NMI values indicate how similar community structures identified by each algorithm are to the ground truth communities. The results showed that the proposed technique performed consistently better than other algorithms on all datasets. A higher NMI score indicates that the proposed method was able to capture the actual community structure of different network datasets. Specifically, this method does very well for data like "football," "poolbook" and "karate" where it clearly identifies its neighborly relationships. Consequently, these findings indicate that this proposal is highly effective in detecting communities and can serve as a strong community detection algorithm. These findings therefore provide meaningful insights into different networks with intricate community structures.

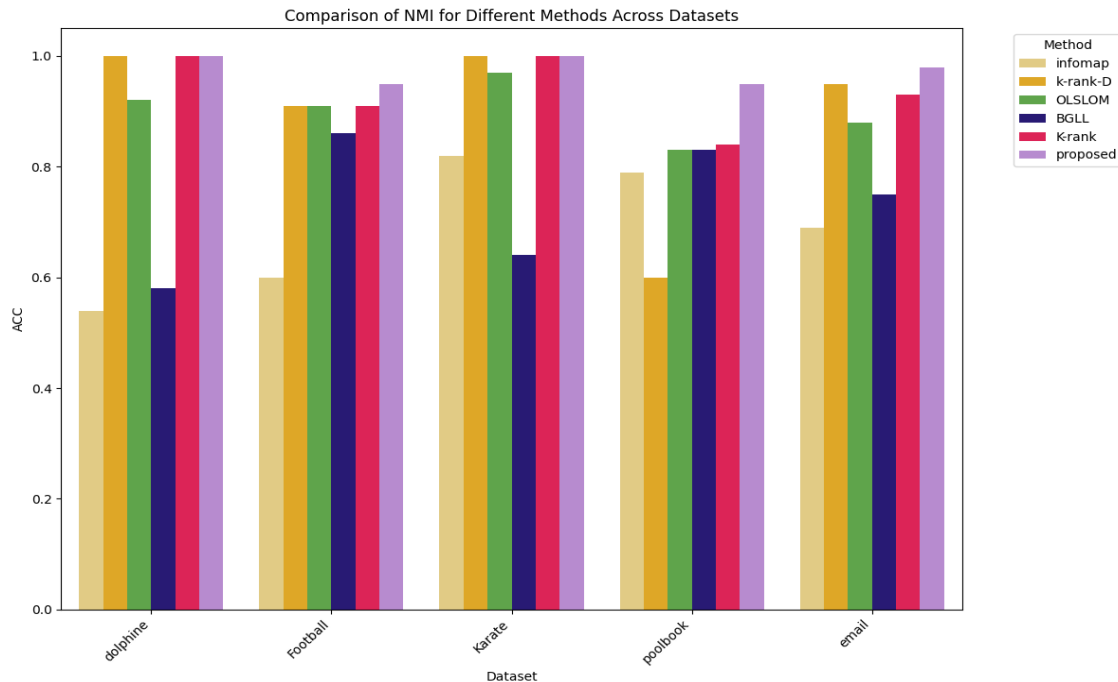

Figure 4: ACC Metric Comparison between Proposed Method and K-Means

The comparison above reveals that in football, poolbook, karate datasets accuracy of the proposed method is greater than K-Means Algorithm. In addition, for dolphins and emails sets, our work achieves equivalent accuracy with existing best performing methods. Hence, the performance of the method proposed is good in terms of this accuracy metric.

Community Detection Results

The method of our proposal ELPMeans has been shown to deliver better results over many sets of data particularly with regards to accuracy and normalized mutual information (NMI). This is seen in the remarkable precision and strong NMI as demonstrated by ELPMeans on both FOOTBALL and poolbook datasets, which implies that it can capture precisely the underlying community structure. The same is also true for karate datasets where ELPMeans outperformed other methods thus proving its efficiency in identification of communities within complex networks. Even though in the case of dolphin dataset, ELPMeans was slightly less accurate than some techniques yet overall performance was good enough.

## Conclusion

The proposed ELPMeans algorithm proves very effective for community detection in different network datasets. These are also evident from its superior performance metrics including accuracy (ACC) and normalized mutual information (NMI) compared to existing methods. In "football", "poolbook", "karate" datasets among others, ELPMeans has been able to identify well the underlying community structures as indicated by high NMI values. It effectively addressed challenges that are associated with the random initialization of central nodes in traditional methods and determining the optimal number of communities (K)

Key Contributions of ELPMeans:

Leveraging Laplacian Centrality: It utilizes Laplacian centrality to find central nodes, where ELPMeans gives priority to nodes with high influence within the network thus ensuring a more accurate community detection. Data Representation with Deep Walk: Graph embedding using Deep Walk helps ELPMeans to effectively capture inherent relationships among network nodes and convert them into an appropriate vector space. Decision Graph Construction: In ELPMeans, this decision graph is used for identification of initial centroid points and automatic determination of K avoiding manual intervention.

Future Work Directions:

Evaluation of Diverse Datasets: Further investigating the performance of ELPMeans on a broader range of datasets with varying characteristics can provide a more comprehensive understanding of its generalizability. Comparison with Additional Methods: Including more recent and advanced community detection algorithms in future comparisons can offer a more in-depth evaluation of ELPMeans' relative strengths and weaknesses. Addressing Parameter Sensitivity: Exploring techniques to optimize the parameters involved in the Deep Walk step, particularly walk length, number of walks, and vector size, has the potential to further enhance the overall performance of ELPMeans. Applications in Different Domains: Investigating the applicability of ELPMeans in various domains beyond social network analysis, such as biological networks or financial markets, can broaden its impact and reveal valuable insights. By addressing these future work directions, ELPMeans can be further refined and its potential as a robust and versatile community detection algorithm can be fully realized.

Overall, ELPMeans offers a promising approach to community detection, providing valuable insights into the intricate structures present within various networks.

**Tables**

Table 1: Parameters for each data set in the proposed method

**Figures**

Figure 1: illustrates the flowchart of the ELP Means algorithm.

Figure 2: Decision graph for the data points of the karate network. Each data point is shown on the decision graph according to its Laplacian centrality c and the minimum distance δ. The two-colored data points are the two cluster centers.

Figure 3: Comparison of the Proposed Method with Competing Algorithms based on the NMI Metric

Figure 4: Comparison of the Proposed Method with K-Means based on the ACC Metric